# A static-loop-current attack against the Kirchhoff-law-Johnson-Noise (KLJN) secure key exchange system

**Mutaz Y. Melhem and Laszlo B. Kish \***

Department of Electrical and Computer Engineering, Texas A & M University, 3128 TAMU,
College Station, TX, 77843, USA; E-Mails: yar111@tamu.edu(MYM); Laszlo.Kish@ece.tamu.edu (L.B.K.)
\*   Correspondence: Laszlokish@tamu.edu; Tel.: +1-979-847-9071

**Abstract:** A new attack against the Kirchhoff-Law-Johnson-Noise (KLJN) key distribution system is explored. The attack is based on utilizing a parasitic dc-voltage-source in the loop. Relevant situations often exist in the low-frequency limit in practical systems, especially when the communication is over a distance, due to a ground loop and/or electromagnetic interference (EMI). Surprisingly, the usual current/voltage comparison based defense method that exposes active attacks or parasitic features (such as wire resistance allowing information leak) does not function here. The attack is successfully demonstrated. Proposed defense methods against it are shown.

## 1. Introduction

*1.1 On secure communications*

Communications systems, standards and technologies have been developed since ancient history. Today we have the internet, Internet-of-Things (IoT), the operating 4th generation wireless networks (LTE), and the expected 5th generation wireless networks. An important requirement of any communication paradigm between these devices is to accomplish the communication securely. That is, to protect the privacy and integrity of the users' data that is transferred over the network. To achieve the security of the transferred data that can contain sensitive information (e.g., bank account credentials, social security number, etc.) it is of utmost importance to defend against attacks. These attacks might be launched by an eavesdropper (Eve) who has access to the information channel between the communicating parties A (Alice) and B (Bob). The attack is passive if it is an eavesdropping without disturbing the channel. The attack is active (invasive) if Eve disturbs or changes the channel, such as during the man-in-the-middle attack. In the present paper, we introduce a new passive attack against the KLJN secure key exchange scheme.

*1.1.1 Secure key exchange*

Secure communication systems employ ciphers to encrypt the message (plaintext) and to decrypt the encrypted message (cyphertext). While the creation of a secure and efficient cipher is a complex problem; this problem is basically solved. The ciphers operate with secure keys that form a momentary shared secret between Alice and Bob. Sharing (exchanging) the key securely is the difficult task. A communicator system cannot be more secure than its key. The security of the key exchange can be conditional or information-theoretic (unconditional).

*1.1.2 Conditional security*

Conditionally secure key exchange systems are the ones used generally nowadays. They are software protocols installed at Alice and Bob. Such algorithms utilize computational complexity and achieve only (computationally-) conditional security, see, e.g [1,2]. The system is temporarily secure provided the adversary has limited computational resources. A major goal of quantum computer developments is to crack these types of key exchange systems (e.g. Shor algorithm). From an *information-theoretic* point of view, the security is non-existent because Eve has all the information to crack the encryption but she needs a long time to do that, unless she has a quantum computer or a yet-to-be-discovered classical algorithm that can do the job in a short time. The security is not future-proof.

*1.1.2 Unconditional (information-theoretic) security*

In order to achieve unconditional (information-theoretic) security at the key exchange, proper laws of physics with a special hardware are utilized. Two major classes of physics-based schemes have emerged for unconditional security:

i) *Quantum key distribution* (QKD) [3,4], concepts assume single photons and utilize quantum physics. The underlying laws of physics are Heisenberg's uncertainty principle and the related quantum no cloning theorem [5]. Even though there are serious debates about the actual level of unconditional security a practical QKD can offer (see, e.g. [6-29]), most scientists agree that QKD is unique by offering information-theoretic security via (a dark) optical fiber, and also through air, during nights, provided the visibility is good.

ii) The *Kirchhoff-law-Johnson-noise* (KLJN) key distribution method that is a based on the statistical physical features of the thermal noise of resistors [30-52]. The related law of physics is the fluctuation-dissipation-theorem (FDT). Some of its advantages are: it works via wire connections including power, phone and internet lines, which can be used as information channel [31, 32] to connect all the homes and the other establishments. It can be integrated on a *chip*, which implies excellent robustness, low price and applicability in bankcards, computers, instruments and *physical unclonable function* (PUF) hardware keys [33, 34]. Its low price allows general applications such as unconditional security for the control of autonomous vehicular networks [35, 36].

*1.2 On the KLJN secure key distribution*

The KLJN scheme [30-52] utilizes the thermal noise of resistors (or the emulation of that by a specific hardware). In the core scheme Alice and Bob have two identical pairs of resistors $R_L$ and $R_H$ ( $R_L < R_H$ ), respectively, see Fig. 1.

The key exchange protocol of a single secure bit is as follows: Alice and Bob randomly pick one of their resistors ( $R_L$ or $R_H$ ), connect it to the wire channel, and keep them there during the bit exchange period while they execute voltage and/or current measurements to learn the resistor value at the other end, see below.

The noise voltage generators shown in Fig. 1 with each resistor can be the resistors' own thermal noise, or an external noise generator emulating a much higher, common *noise-temperature* that is publicly agreed. The power density spectra of the voltage and current in the channel are given by the by the Johnson-Nyquist formulas [11] :

$$S_u(f) = \frac{4kTR_A R_B}{R_A + R_B} \quad (1)$$

$$S_i(f) = \frac{4kT}{R_A + R_B} \quad , \quad (2)$$

where $k$ is the Boltzmann's constant, $T$ is the common temperature, $R_A$ and $R_B$ are the actually connected resistances at Alice's and Bob's ends, respectively, $R_A, R_B \in \{R_L, R_H\}$. After the measurement and spectral analysis, Equations (1) and (2) have two unknown variables, the values of $R_A$ and $R_B$ , thus Eve can find the values of the connected resistors, but no necessarily their locations, by solving these equations.

We can represent the 4 different situations of the connected resistors ( $R_L$ and/or $R_H$ ) at Alice's and Bob's ends by the indices of the connected resistors, LL, LH, HL, and HH, respectively. As all the resistors have the same (noise) temperature, the ideal system is in thermal equilibrium where the 2nd law of thermodynamics guarantees zero net power-flow. Hence, Eve cannot use the evaluation of power flow to determine the locations of the momentarily connected resistors unless they have the same resistance values. On the other hand, Alice and Bob can determine the connected resistor values by Equations (1) or (2) since they know the value of their own connected resistors. When $R_A = R_B$, which happens at 50% of the bit exchange attempts, the results are discarded.

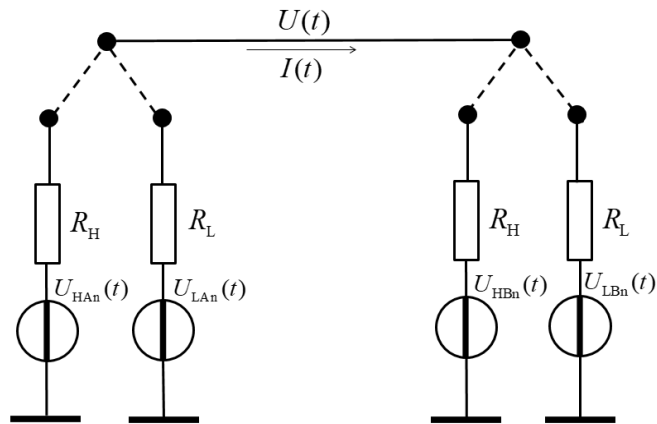

**Figure 1.** The core of the KLJN system. $U_{HAn}(t)$, $U_{LAn}(t)$, $U_{HBn}(t)$ and $U_{LBn}(t)$ are the (thermal) noise voltage generators for the related resistances. $U(t)$ and $I(t)$ are the measured noise voltage and the current in the wire that are used to evaluate the power density spectra $S_u(f)$ and $S_i(f)$, respectively.

*1.2.1 On former attacks against the KLJN secure key distribution*

Several attacks have been proposed but no attack has been able to compromise the unconditional security of the KLJN scheme because each known attack can efficiently be nullified by a corresponding defense scheme.

The attacks can be categorized into two classes:

i) *Passive attacks* that utilize the non-ideal or parasitic features in a practical KLJN system for information leak. Non-zero wire resistance, see in [37], [38] poses the most known threat and the most efficient attack is power balance measurement (Second Law Attack) [39]. An efficient defense is based on a proper temperature-offset [39, 40]. Temperature-inaccuracies [41] and resistance-inaccuracies [42] can also cause information leak. On the other hand, these inaccuracies can compensate each other out [43] if used in a creative way. Non-zero cable capacitance [44] or cable inductance can also yield information leak that can be fixed by specific design including the proper choice of the frequency range and privacy amplification. Transients can also be utilized for attack [45] but there are various ways of defenses [46, 47]. The newest KLJN system, the random-resistor-random-temperature (RRRT-KLJN) scheme [48] is robust against the above vulnerabilities; at least, no known attack exists against it yet.

ii) *Active attacks*, where Eve either modifies the information channel or she injects an extra current into that. Current injection attack [30, 49] and the man-in-the-attack [50] are the explored examples [2006]. Due to the current and voltage comparison [50] feature and its more advanced cable-modeling version [49], active attacks are, so far, the least efficient attacks against the KLJN scheme.

iii) *Flawed attacks*. There are some proposed attack methods that are based on misconceptions and they do not work. See their brief summary and their criticism, for example, in papers [51-55] and book [56].

## 2. The new attack scheme utilizing deterministic loop currents

*2.1 The situation that Eve's utilizes for the attack*

In practical KLJN systems, in order to save a wire, the common end of the resistors (see Fig. 1) is often connected to the ground. At practical situations, there is often an imbalance, a voltage difference between various locations of the ground; for example due to ground loop currents or electromagnetic interference (EMI) [53]. This potential information leak was pointed out in [53] as the potential source of information leak in the case of significant cable resistance. However, it has not been realized in [53] that information leak can exist even at zero cable resistance.

In this paper, we explore this new information leak in the DC parasitic voltage limit. Thus the considerations hold for situations where during the bit exchange period, the relative change of the parasitic voltage is small. For the sake of simplicity but without the limitation of generality, we assume that the imperfection is represented by a *positive* DC voltage generator located at Alice's end, see Fig. 2.

Due to Kerckhoffs's principle of security, that is, the assumption that the enemy knows everything except momentary key, we must assume that Eve knows the polarity and value of this DC voltage. (If she does not know it at the beginning, she will be able to extract it by a long-time averaging). The direction of the current $I(t)$ is assumed to point from Alice to Bob. The voltage $U(t)$ and current $I(t)$ in the wire contain the sum of a DC and an AC (stochastic, that is, noise) components.

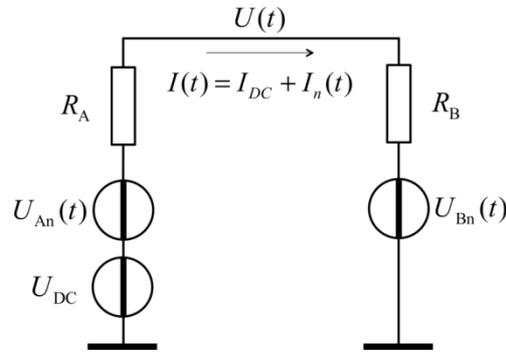

**Figure 2.** The KLJN system with ground loop voltage. Here $U_{An} \in \{U_{LAn}; U_{HAn}\}$ and $U_{Bn} \in \{U_{LBn}; U_{HBn}\}$, are the voltage noises belonging to the randomly chosen resistors, $R_A \& R_B \in \{R_L; R_H\}$, of Alice and Bob, respectively. $U_{DC}$ is the ground loop DC voltage source and $U(t)$ & $I(t)$ are the voltage and current on the wire, respectively.

Let us analyze the resulting voltages and currents. The current in the wire is:

$$I(t) = I_{DC} + I_n(t) \tag{3}$$

where $I_{DC}$ is its DC component

$$I_{DC} = \frac{U_{DC}}{R_A + R_B} \tag{4}$$

and $I_n(t)$ is its AC (noise) component

$$I_n(t) = \frac{U_{An}(t) - U_{Bn}(t)}{R_A + R_B}, \qquad (5)$$

where $U_{An}$ and $U_{Bn}$, with $U_{An} \in \{U_{LAn}; U_{HAn}\}$ and $U_{Bn} \in \{U_{LBn}; U_{HBn}\}$, are the voltage noise sources of the chosen resistors, $R_A$ and $R_B$, respectively.

The voltage on the wire is:

$$U(t) = I(t) R_B + U_{Bn}(t) \qquad (6)$$

From Equations 3 and 6 we obtain

$$U(t) = U_{DCw} + U_{ACw}(t) = I_{DC} R_B + I_n(t) R_B + U_{Bn}(t) \qquad (7)$$

where $U_{DCw}$ and $U_{ACw}(t)$ represent the DC and AC voltage components in the wire, respectively. The DC component can be written as:

$$U_{DCw} = I_{DC} R_B = \frac{U_{DC}}{R_A + R_B} R_B \qquad (8)$$

The DC component is different during Alice's and Bob's LH and HL bit situations of secure bit exchange, which yields information leak. In the LH situation, that is, when $R_A = R_L$ and $R_B = R_H$, the DC component of the voltage on the wire is

$$U_{DCw} \equiv U_{LH} = U_{DC} \frac{R_H}{R_H + R_L}, \qquad (9)$$

and, in the HL bit situation:

$$U_{DCw} \equiv U_{HL} = U_{DC} \frac{R_L}{R_H + R_L}. \qquad (10)$$

Note, as we have been assuming that in the given KLJN setup $R_H > R_L$, in this particular situation

$$U_{HL} < U_{LH} \qquad (11)$$

For later usage, we evaluate the average of $U_{LH}$ and $U_{HL}$, and call this quantity *threshold voltage*, $U_{th}$, where:

$$U_{th} = \frac{U_{LH} + U_{HL}}{2} = \frac{U_{DC}}{2} \tag{12}$$

The effective (RMS) amplitude $U_{ACw}$ of the noise voltage on the wire is identical in both the LH and HL cases:

$$U_{ACw} = \sqrt{4kTB_W \frac{R_L R_H}{R_L + R_H}} \ . \tag{13}$$

Note, the voltage and current noises in the wire follow normal distribution since the addition of normally distributed signals result in a signal that has normal (Gaussian) distribution with a corresponding mean (see Equation 10) and variance.

For an illustration of the information leak, see Figure 3. The DC component, that is, the mean value of the resulting (AC+DC) Gaussian depends on the bit situation during secure key exchange. This dependence poses as a source of information for Eve about the secret key. This feature will be exploited below for the new attack scheme.

*2.2 The attack scheme*

The attack consists of three steps: measurement, evaluation, and guessing.

*i) Measurement*: During a single secure bit exchange, Eve measures $N$ independent samples of the wire voltage.

*ii) Evaluation*: She evaluates the fraction $\gamma$ of these $N$ samples that are above $U_{th}$:

$$\gamma = \frac{N^+}{N} \ , \tag{14}$$

where $N^+$ is the number of samples that are above $U_{th}$.

*iii) Guessing* (based on Equations 9-14): For $0.5 < \gamma$ and $\gamma < 0.5$, Eve's guesses are the LH and HL bit situations, respectively. For $\gamma = 0.5$ her decision is undetermined and carries no useful information.

*iv)* Eve's correct guessing probability $p$ is given as:

$$p = \lim_{n_{tot} \to \infty} \frac{n_{cor}}{n_{tot}} \ , \tag{15}$$

where $n_{tot}$ is the total number of guess bits and $n_{cor}$ is the number of correctly guessed bits. The situation $p = 0.5$ indicates perfect security against Eve's attack.

In the next section, we demonstrate the attack method by computer simulations.

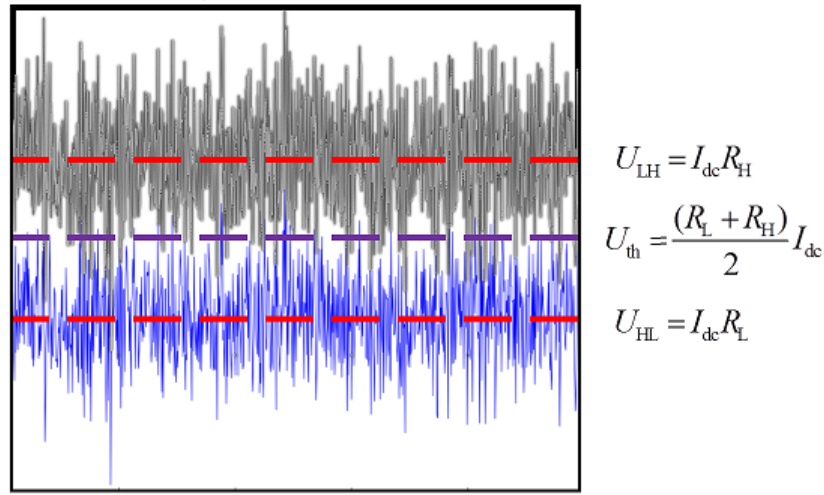

**Figure 3**. Eves' threshold scheme to guess the bit situation LH *vs* HL.

## 3. Simulation Results

To test Eve's correct guessing probability $p$ for the LH situation. We assumed that Alice and Bob selected $R_L = 1$ k$\Omega$ and $R_H = 10$ k$\Omega$. During these experiments, the DC voltage was kept at a constant level of 0.1 V (see Figures 2 and 3). To generate the noise, we used the white Gaussian noise function (wgn) from the communication system toolbox of Matlab to test the success statistics of the attack scheme while varying the temperature. The effective bandwidth $\Delta f$ and the range of temperatures were 1 MHz and $10^8 < T < 10^{18}$ K, respectively. At the lower temperatures $p$ was 1 within the statistical inaccuracy of simulations and at the high-temperature limit it converged to 0.5. The duration of the *secure bit exchange period* was characterized by the number $N$ of *independent* noise samples used during the exchange of the particular bit.

We tested secure key length $M$ =700 bits at different bit exchange durations represented by sample/bit numbers $N$ = 1000, 500 and 200, respectively. Figure 4 shows Eve's correct guessing probability ($p$) of a key bit vs. temperature. With temperature approaching infinity, the effective noise voltage on the wire is also approaching infinity and the Gaussian density function will be symmetrically distributed around the threshold voltage $U_{th}$. Thus the probabilities of finding the noise amplitude above or below $U_{th}$ are identical (0.5) Then Eve's correct guessing probability of represents the perfect security limit, $p = 0.5$.

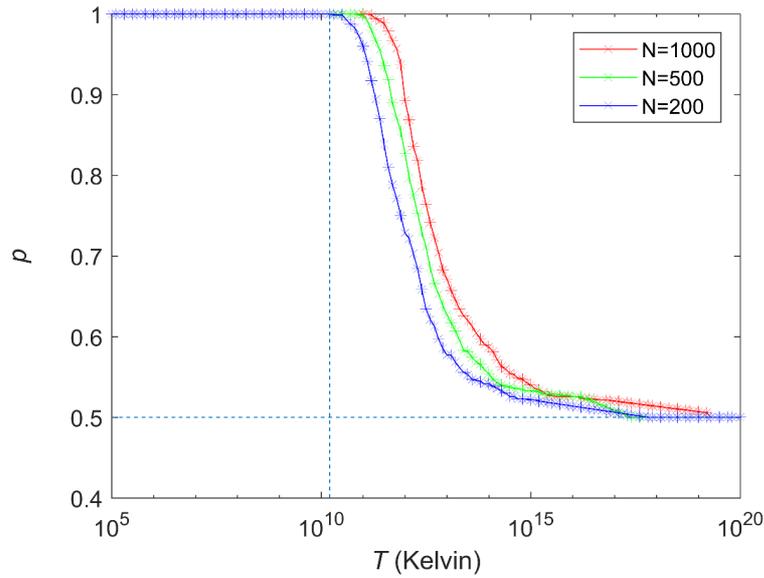

**Figure 4.** Eve's correct guessing probability (*p*) of key bits vs. temperature at bandwidth $\Delta f = 10^6$ Hz, for key length 700 bits, and duration/bit (number of samples/bit) 200, 500 and 1000, respectively. The limit *p*=0.5 stands for perfect security.

The observed dependence can be interpreted by the behavior of the error function (see also Equations 8 and 12):

$$p\{U(t) \geq U_{th}\} = 0.5\left[1 - erf\left(\frac{U_{th} - U_{DCw}}{U_{eff}}\right)\right] \quad , \quad (16)$$

where $U(t)$ is the instantaneous voltage amplitude in the wire and the error function is:

$$erf(x) = \frac{1}{\sqrt{\pi}} \int_{-x}^{x} \exp^{-y^2} dy \quad (17)$$

The noise in the KLJN scheme is a bandlimited white noise thus, in accordance with the Johnson formula, the effective noise voltage scales as:

$$U_{eff} \propto \sqrt{T \Delta f} \quad . \quad (18)$$

Therefore, when the temperature T is converging to infinity, *p* is converging to the perfect security limit of 0.5, see Figure 4.

## 4. Some of the possible defense techniques against the attack

Based on the considerations above, the impact of the attack can be eliminated by various means. The most natural ways are:

i) Cancelling the effect of the DC-Voltage sources. For example, Bob can put a variable DC source that compensates its effect. Similarly, eliminating ground loops is also beneficial.

ii) Alice and Bob can increase the effective temperature, that is the amplitudes of their noise generators, see Equation 18 and Figure 4.

iii) Alice and Bob can increase the bandwidth to increase the effective value of the noise, see Equations 18 and 20. However, the bandwidth must stay below the wave limit [54] to avoid information leak due to reflection thus the applicability of this tool is strongly limited.

## 5. Conclusion

The KLJN secure key exchange scheme is a statistical physical system that offers unconditional (information-theoretic) security. For a detailed survey and its history, see a recent book [56].

In this paper a novel attack against the KLJN protocol is shown that uses a frequently occurring parasitic feature, namely the imbalance of voltages between the ground points at the two ends. We showed that such parasite voltages and currents could cause information leak. The attack was demonstrated by computer simulations and proper defense protocols were shown to eliminate the information leak. The considerations in this introductory paper about this new attack type are valid in the limit when the parasitic voltage is static or its (possible) time-dependence is slow compared to the dynamics of the noise. Work is in hand to explore more complex situations, too.

13 of 1353. Chen, HP, LB Kish, CG Granqvist. "On the "Cracking" Scheme in the Paper "a Directional Coupler Attack against the Kish Key Distribution System" by Gunn, Allison and Abbott." *Metrology and Measurement Systems* 21 (2014): 389-400.

54. Chen, HP, LB Kish, CG Granqvist, G Schmera. "Do Electromagnetic Waves Exist in a Short Cable at Low Frequencies? What Does Physics Say?" *Fluct. Noise Lett.* 13 (2014): 1450016.

55. Kish, LB, Z Gingl, R Mingesz, G Vadai, J Smulko, CG Granqvist. "Analysis of an attenuator artifact in an experimental attack by Gunn-Allison-Abbott against the Kirchhoff-law-Johnson-noise (KLJN) secure key exchange system." *Fluct. Noise Lett*. 14 (2015): 1550011.

56. Kish, LB. "The Kish Cypher. The Story of KLJN for Unconditional Security." World Scientific (2017). https://doi.org/10.1142/8707